\documentclass[twocolumn, floatfix]{revtex4-2}
\usepackage{braket}
\usepackage{amsmath}
\usepackage{amssymb}
\usepackage{amsfonts}
\usepackage{multirow}
\usepackage{epstopdf}
\usepackage{subcaption}
\usepackage{bbm}
\usepackage{graphicx, epstopdf}
\DeclareGraphicsExtensions{.eps}
\begin{document}

\title{Optimizing the Phase Estimation Algorithm \\ Applied to the 
  Quantum Simulation \\ of Heisenberg-Type Hamiltonians
}
\author{Scott Johnstun}
\email{scottjohnstun@byu.net}
\affiliation{Department of Physics and Astronomy, Brigham Young University}

\author{Jean-Fran\c{c}ois Van Huele}
\email{vanhuele@byu.edu}
\affiliation{Department of Physics and Astronomy, Brigham Young University}
\date{\today}

\begin{abstract} The phase estimation algorithm is a powerful quantum algorithm with applications in cryptography, number theory, and simulation of quantum systems. We use this algorithm to simulate the time evolution of a system of two spin-1/2 particles under a Heisenberg Hamiltonian. The evolution is performed through both classical simulations of quantum computers and real quantum computers via IBM's Qiskit platform. We also introduce three optimizations to the algorithm: circular, iterative, and Bayesian. We apply these optimizations to our simulations and investigate how the performance improves. We also discuss the paradigms of iterative and update-based algorithms, which are attributes of these optimizations that can improve quantum algorithms generally. 
\end{abstract}

\maketitle

\section{Background}


Since the turn of the 21st century, quantum information has become nothing short of a revolution in physics. Quantum information exists as a union between quantum mechanics, statistical physics, and information theory and has recently acquired a mainstream following with the promising prospect of quantum computing. Richard Feynman first proposed quantum systems as a computational tool in 1985 \cite{feynman_qc, hey2018feynman}. Since then, algorithms have been found that are expected to, in theory, solve difficult problems faster than any classical computer possibly could \cite{quantum_algorithms}. These include Peter Shor's algorithms for factoring large integers and computing discrete logs \cite{shor}; Grover's algorithm for efficient database search \cite{grover}; and various algorithms for simulation of quantum systems \cite{quantum_simulation}. With these (and other) algorithms and sufficiently capable quantum computers, many difficult or unsolved problems in mathematics, physics, chemistry, and other fields could be solved efficiently. 

Requisite for the implementation of such algorithms is the development of physical systems that can function as large, controllable quantum computers with minimal noise. Quantum computers are made up of a collection of ``qubits''\index{Qubit}, the quantum version of classical bits with which classical computers compute. Whereas a classical bit can only hold a value of 0 or 1, a qubit is a quantum system that can hold any normalized linear combination of two states referred to as $\ket{0}$ and $\ket{1}$. Similar to bits in a classical computer, qubits serve as both computational space and storage space for quantum computers. However, whereas modern classical computers contain billions of bits with which to compute, modern quantum computers have less to work with: at the end of 2020, the largest quantum computers to date were a 76-qubit photonic quantum computer produced by a team of researchers in China \cite{big_computer} and a 72-qubit superconducting quantum computer produced by Google \cite{google_computer}. All quantum computers are also susceptible to noise altering the state of their qubits. Noise will be described more specifically, and in the context of our work, in Section \ref{sec:IBMQ}. 

\section{Quantum Simulation}
We are particularly interested in the use of quantum computers for quantum simulation\index{Quantum simulation}, which is the process of using controllable quantum systems to simulate other, harder-to-control systems. Quantum simulation has promising applications in developing new drugs, making sense of certain chemical reactions, and creating new materials \cite{waiting_for_quantum_simulation}. This is because classical simulation of quantum systems is inherently inefficient. The amount of memory required to represent a quantum system on a classical computer grows exponentially as the size (number of particles, degrees of freedom, etc.) increases: a system of 40 spin-1/2 particles would require 4 terabytes to represent classically \cite{quantum_simulation}. On the other hand, a quantum simulator would only require 40 qubits plus perhaps a few extra qubits, depending on the objective of the simulation. Evidently classical simulation algorithms for quantum systems are naturally limited to small systems, prompting the need for quantum simulation algorithms for future scientific discovery. 

Quantum simulation can be divided into digital and analog quantum simulation\index{Quantum simulation!digital vs. analog}. In analog quantum simulation, we have two systems, one controllable and one difficult to control, and a mapping between the two systems such that evolution of the controllable system corresponds to evolution in the difficult system. Analog quantum simulation is currently more realistic than digital quantum simulation, yet examples of systems and mappings between them are rare \cite{analog_sim}. Digital quantum simulation, on the other hand, uses programmable universal quantum computers---quantum computers on which any unitary operation can be implemented---to simulate quantum systems. Digital quantum simulation is more versatile, allowing many types of systems to be simulated with a single physical system \cite{Lanyon57}, yet currently universal quantum computers are both too small to simulate systems beyond those that can be simulated classically and too noisy to produce reliable results.

In this work we use digital quantum simulation\index{Quantum simulation!of Heisenberg Hamiltonian} to evolve a system of two spin-1/2 particles described by a Heisenberg Hamiltonian and recover its first excited energy. This problem is entirely solvable on a classical computer (or with analytic methods), but our implementation of a simulation algorithm and its optimizations demonstrates some of the same challenges and questions that must be addressed in order for a large scale quantum computer to simulate the evolution of more complex systems. We assume that the reader has some basic familiarity with quantum algorithms and notation; for general references on the subject, see \cite{mikenike} and \cite{mermin}.

\section{Quantum Phase Estimation Algorithm}
Among the most celebrated algorithms in quantum computing is the Phase Estimation Algorithm (PEA)\index{Phase estimation algorithm}. The purpose of the PEA is to estimate the phase $\phi$ conferred by a unitary operator $U$ to an eigenstate $\ket{u}$. Mathematically, if $\ket{u}$ is an eigenstate of $U$, then
\[
    U\ket{u} = \lambda \ket{u},
\]
which indicates that $\lambda$ is the eigenvalue corresponding to the eigenstate $\ket{u}$. If $U$ is unitary, then $\lambda$ must satisfy the condition $|\lambda|^2 = 1$; since $\lambda$ may be complex, we can write $\lambda = e^{2\pi i \phi}$ for some $\phi \in [0,1)$. The goal of the PEA is to estimate $\phi$ given $U$. One common usage of unitary operators is for time evolution operators in quantum mechanics. We will decide on a choice for $U$ in Section \ref{section:heisenberg}. As we will see, knowing $\phi$ allows us to estimate the energy of a system evolving in time, since the rate at which the phase of a stationary state oscillates is proportional to its energy.

The PEA can be represented on a quantum circuit\index{Quantum circuit} with two quantum registers. The first register is the \textit{counting} register, which can hold any amount of qubits; we denote the number of qubits by $n$. This register is the register that will actually store the estimate of the phase, so more qubits in this register allows for a more accurate estimate of the phase. The second register is known as the \textit{state} register, and it exists solely to hold the eigenstate $\ket{u}$ from which the phase is deduced. The PEA performs its task in three steps: superposition, wherein the two registers are initialized into appropriate states; phase kickback, wherein simulation happens and phase information is recorded; and Quantum Fourier Transform (QFT)\index{Quantum Fourier transform} (see Chapter 5 of \cite{mikenike}), wherein the recorded phase information is converted to a binary representation through the inverse QFT. Note that in the following discussion, a subscript after a state indicates the number of qubits that state represents, and for an $m$-bit integer $x$, $\ket{x}_m$ represents the $m$-qubit state $\ket{x_m}\ket{x_{m-1}}\dots \ket{x_1} \ket{x_0}$, where $x_i$ is the $i^{th}$ bit of the binary representation of $x$. Tensor products, represented by $\otimes$, between the kets are implied; as an exponent to an operator or state $U$, we denote $U^{\otimes n} \equiv U \otimes U \otimes \dots \otimes U$, where $n$ instances of $U$ are linked by $n-1$ tensor products. The three steps of the PEA are:
\begin{enumerate}
    \item \textit{Superposition}: Initialize a uniform superposition of all $2^n$ possible states on the $n$-qubit counting register:
    \[
        \ket{0}_n\;\; \longrightarrow \;\; H^{\otimes n} \ket{0}_n = \frac{1}{2^{n/2}} \left( \ket{0} + \ket{1} \right)^{\otimes n},
    \]
    where $H$ is the Hadamard gate. If necessary, also initialize the state register to the eigenstate $\ket{u}$ (here assumed to be known beforehand; see \cite{circular_optimization} for PEA with superposition states):
    \[
        \ket{0}_m \;\; \longrightarrow \;\; \ \ket{u},
    \]
    where $m$ is the number of qubits required to represent $\ket{u}$; we have dropped the $m$ subscript from $\ket{u}$ itself for brevity.
    \item \textit{Phase Kickback}: For each $k$ from 0 to $n-1$, use the $k^{th}$ qubit of the counting register as the control for a controlled-$U^{2^k}$ gate acting on the state register. Each application performs the following transformation on the partial system consisting of only the $k^{th}$ qubit of the counting register and the full state register
    \[
        \frac{1}{\sqrt{2}}(\ket{0} + \ket{1}) \ket{u} \;\; \longrightarrow \;\; \frac{1}{\sqrt{2}} (\ket{0} + e^{2 \pi i \phi \cdot 2^k} \ket{1})\ket{u}
    \]
    since $U\ket{u} = e^{2 \pi i \phi}\ket{u} \implies U^{2^k} \ket{u} = e^{2 \pi i \phi 2^k}\ket{u}$ and the controlled nature of the gate ensures $U$ is only applied in the subspace where the $k^{th}$ counting qubit is in the $\ket{1}$ state. Note that since $\ket{u}$ is an eigenstate of $U$ and the phase transfers to the counting register, the state register remains unchanged. That is why this step is referred to as phase kickback\index{Phase kickback}.
    Once this has been applied to all $n$ register qubits, the counting register is left in the state
    \begin{equation}\label{eqn:iqft}
        \ket{count}_n = \frac{1}{2^{n/2}} \sum_{k=0}^{2^n - 1} e^{2 \pi i \phi k} \ket{k}_n.
    \end{equation}
    \item \textit{Quantum Fourier Transform}: The right hand side of Eq. (\ref{eqn:iqft}) happens to be exactly the QFT of input $\ket{2^n \phi}_n$, the state representing the number $2^n \phi$ (see Section 5.2 of \cite{mikenike}). Therefore, applying the inverse quantum Fourier Transform (denoted as $QFT^\dag$, since the QFT is unitary) to the counting register will leave it in the state
    \[
        QFT^\dag (QFT\ket{2^n \phi}_n) = \ket{2^n \phi}_n,
    \]
    from which phase information can be obtained. We will detail specific methods for obtaining the phase as well as variations of this algorithm in Section \ref{sec:implementation}.
\end{enumerate}

\section{Heisenberg Hamiltonian}\label{section:heisenberg}
In the discussion of the implementation of the PEA, we did not specify $U$. In this work we choose to use a Heisenberg Hamiltonian\index{Heisenberg Hamiltonian}, which represents a spin interaction between two particles
\begin{equation}\label{eqn:hamiltonian}
    H = \frac{\lambda}{4} (\boldsymbol{\sigma}_1 \cdot \boldsymbol{\sigma}_2) = \frac{\lambda}{4} (X_1 X_2 + Y_1 Y_2 + Z_1 Z_2),
\end{equation}
where $X_i$, $Y_i$, and $Z_i$ are the Pauli spin operators\index{Pauli matrices} corresponding to particle $i$ and $\boldsymbol{\sigma}_i = X_i \hat{x} + Y_i \hat{y} + Z_i \hat{z}$. This Hamiltonian is parameterized by $\lambda$, which controls the strength of the interaction. 

Evolution under this Hamiltonian is governed by its time evolution operator 
\begin{equation}\label{eqn:time_evolution}
    U = \exp{(-i H t/ \hslash )} = \exp{(-i H \tau)} \equiv U(\tau),
\end{equation}
where we have defined the evolution parameter $\tau = t/\hslash$.\index{Evolution parameter $\tau$} Note that $\tau$ has units of inverse energy but increases exactly proportionally to time during the simulation, so it can be thought of as a parameter akin to time. Applying $U$ to an eigenstate yields the result
\[
    U \ket{u} = \exp(-i H \tau) \ket{u} = \exp(-i \epsilon \tau) \ket{u}.  
\]
Here $\epsilon$ denotes the system's energy. Estimating the phase produced by the Heisenberg Hamiltonian for various values of $\tau$ gives a function $\phi(\tau)$ which satisfies
\[
    2\pi i \phi(\tau) = -i \epsilon \tau.
\]
Differentiating with respect to $\tau$ and solving for $\epsilon$ gives us the important equation
\begin{equation}\label{eqn:energy}
    \epsilon = -2\pi \frac{d\phi}{d\tau}.
\end{equation}
This is how phase estimation can be used to determine energy levels in a system.

\begin{figure}[tb]
    \centering
    \includegraphics[scale=0.58]{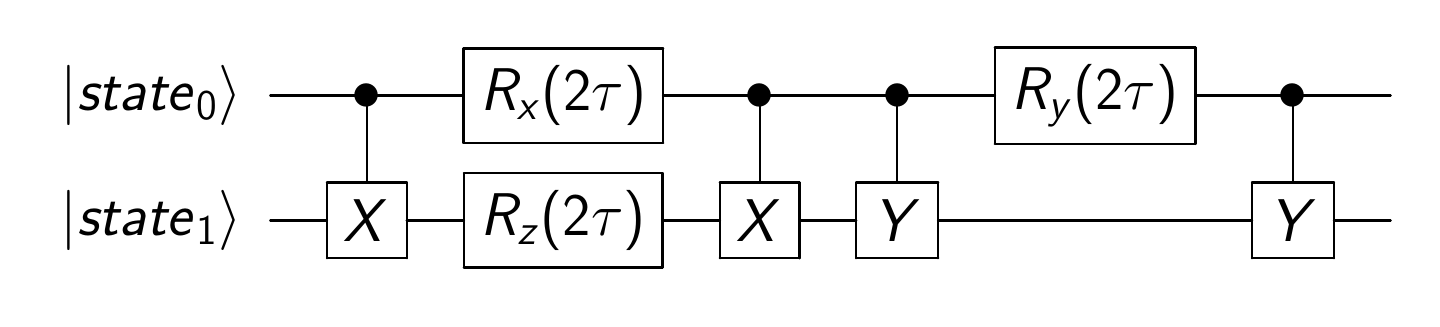}
    \caption[Quantum circuit diagram for time evolution operator of Heisenberg Hamiltonian]{Quantum circuit\index{Quantum circuit} diagram for the Heisenberg Hamiltonian's time evolution operator $U(\tau)$ in Eq. (\ref{eqn:time_evolution}). The $X$ and $Y$ gates indicates Pauli $X$ and $Y$ operators respectively. The $R_x$, $R_y$, and $R_z$ gates indicate a Pauli rotation about the corresponding axis by the input angle $2\tau$. To implement a controlled $U$ gate, only the three rotation gates need to be controlled by a qubit external to the system.}
    \label{fig:Ucircuit}
\end{figure}

Implementing $U$ on  a quantum computer is a nontrivial task. Although any unitary operator can be represented on a quantum computer, decomposing one into a sequence of simple unitary operations that can be implemented on a quantum computer is difficult. In Cruz \textit{et al.} it was shown how the time evolution operator corresponding to a Hamiltonian containing $Z_1 Z_2$ can be decomposed into $cNOT$ gates and a $Z$ rotation gate \cite{circular_optimization}. Following this result, we implement the Heisenberg Hamiltonian \index{Heisenberg Hamiltonian} with the gate sequence pictured in Fig. \ref{fig:Ucircuit}. Gates involving $X$, $Y$, or $Z$ refer to the corresponding Pauli operators. A filled circle on a qubit indicates that it is the control for the operation it is linked to. If $T$ is an $n$-qubit gate, the controlled-$T$ gate is the $(n+1)$-qubit gate defined by $\ket{0}\bra{0}\otimes \mathbbm{1} + \ket{1}\bra{1} \otimes T$. \index{Controlled gate} Armed with this decomposition, we are able to use the PEA to simulate evolution under the Heisenberg Hamiltonian and determine energy levels. 



\section{IBM Q System}\label{sec:IBMQ}
The IBM Q System provides free public access to superconducting quantum computers maintained by IBM \cite{IBMQ}, which we use to run our experiments. Experiments on these computers can be designed in Qiskit\cite{Qiskit}, a Python library maintained by IBM, then sent to a quantum computer to be executed. In this way, we can run experiments on actual quantum computers to determine how well our algorithms perform in real conditions. Qiskit also provides functionality to simulate the result of a computation with classical computers among its many features. Running an experiment on an IBM quantum computer amounts to creating a circuit to be run, specifying the amount of shots (times the computer is initialized, the circuit is run, and the qubits are measured; multiple shots are necessary to reduce uncertainty due to the probabilistic nature of measurement outcomes), sending the experiment request, and obtaining measurement results. 

Noisy Intermediate Scale Quantum (NISQ) computers---including those available through IBM Q---are prone to challenges that classical computers do not have to deal with. Having far more states available to them, qubits are more sensitive to small perturbations than classical bits are. Qubits\index{Qubit!superconducting} in superconducting quantum computers like those available through IBM are affected by three types of stochastic noise \index{Noise}\cite{youtube}: Johnson noise (or Johnson-Nyquist noise), which is white noise proportional to temperature; quantum noise, which is noise that results from the possibility of the qubit releasing energy to its environment, even at zero temperature; and classical 1/f noise from stray electromagnetic fields, which can cause dephasing to occur. In addition to these stochastic errors, systematic errors from imperfect hardware or incorrect pulse timing may also affect the computers \cite{superconducting_noise}. Larger circuits require more time to run, making them more prone to errors than shorter circuits\index{Error}. In this context, circuit length is determined by the number of gates. In the following sections we will see how circuit length affects the success rate of each algorithm.

\onecolumngrid
\begin{figure*}[tb]
    \centering
    \includegraphics[scale=0.80]{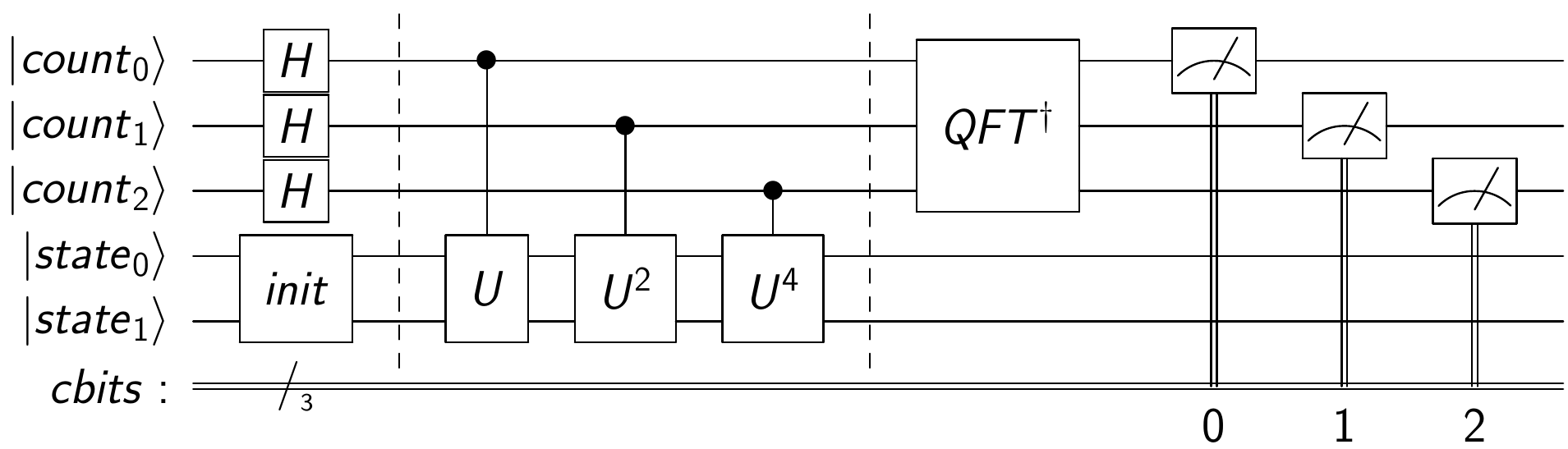}
    \caption{Circuit diagram for the base implementation of the PEA with a Heisenberg Hamiltonian on a 5-qubit quantum computer. The upper three qubits form the counting register and the lower two form the state register. The double line at the bottom represents the three classical bits (indicated by the slash with a three) onto which measurement information is stored. Numbers below measurements indicate the classical bit onto which the measurement result is read.}
    \label{fig:base}
\end{figure*}

\twocolumngrid

\section{Implementation of the Phase Estimation Algorithm}\label{sec:implementation}
In this section we describe our four algorithms for implementing the PEA. Throughout this section $\phi$ refers to the true phase and $\hat{\phi}$ refers to an estimate of the phase produced by an algorithm.

\subsection{Base Algorithm}

We first detail our base implementation of the PEA\index{Phase estimation algorithm} on IBM quantum computers. By base implementation we refer to the common implementation of the PEA as found in textbooks on quantum computing \cite{mikenike, Qiskit-Textbook}. Most public quantum computers at IBM have five qubits\index{IBM Q}, so we choose to implement a five-qubit version of the PEA. We also use three classical bits to record information onto. The resulting circuit is pictured in Fig. \ref{fig:base}. Since our Hamiltonian acts on two qubits\index{Qubit}, we choose to use two qubits for our state register. This leaves three qubits for the counting register. Three Hadamard gates on the input register put it into the desired superposition state while the $init$ gate initializes the state register to the eigenstate $(\ket{01} + \ket{10})/\sqrt{2}$ of the Heisenberg Hamiltonian. This choice of eigenstate is arbitrary. We then apply a sequence of controlled $U$ gates, where $U$ is the time evolution operator corresponding to the Heisenberg Hamiltonian, acting on the state register and controlled by a single qubit on the counting register. The $k^{\text{th}}$ qubit on the counting register controls the operation $U^{2^k}$. This sequence results in a counting register state of $QFT(\ket{2^3 \hat{\phi}})$, so we apply the inverse Quantum Fourier Transform\index{Quantum Fourier transform} $QFT^\dag$ to recover the state $\ket{2^3 \hat{\phi}}$. Finally, we measure each of the counting qubits and record the result on the corresponding classical bit. After performing many shots of this circuit, we obtain a distribution of binary integers betwen zero and seven, inclusive, which we represent as $P(x) :  \{0, 1, ..., 7\} \rightarrow \mathbb{R}$. For the base implementation of the PEA, we simply choose the result that happened with the largest frequency, $\underset{x}{\text{argmax}}\; P(x)$. Dividing this result by eight reveals an approximation of the phase:
\[
    \hat{\phi}_{maj} = \frac{\underset{x}{\text{argmax}}\; P(x)}{8}.
\]
This technique is simple, but it is limited in resolution. In general, we can obtain one binary digit of precision per counting qubit. With only three counting qubits, the estimator $\hat{\phi}$ is limited to the eight possibilities $\{k/8\}_{k=0}^7$, whereas $\phi$ itself can be any real number in the interval $[0, 1)$. This leads to discontinuous jumps in $\hat{\phi}$ as $\phi$ varies. Thus the base algorithm computes a rudimentary phase estimate that is limited in precision.

\subsection{Circular Optimization}
Circular statistics\index{Circular statistics} offer one way to optimize the PEA. This first optimization uses the same circuit as the base PEA implementation, pictured in Fig. \ref{fig:base}. The optimization consists of how we process the measurement data. Instead of using a simple majority rule, we calculate the circular mean of the data and use its argument as the phase estimator. Given a measurement probability distribution $P(x)$ defined for $N$ integers $x$, its circular mean $\mu$ is defined as
\[
    \mu = \sum_{x=0}^{N-1} P(x) e^{i \pi x/N}.
\]
The circular mean is a complex number whose argument can be used as an approximator for $\phi$:
\[
    \hat{\phi}_{circ} = \frac{1}{2\pi} \arg \mu.
\]
This is a more natural estimator for $\phi$ than the majority rule because, like $\phi$, $\arg \mu$ is a circular measure. In addition to this, utilizing the circular mean in this way allows for better resolution in $\hat{\phi}$. Although $\hat{\phi}$ cannot always exactly represent $\phi$, it does vary smoothly as $\phi$ changes. Cruz \textit{et al.} showed that the circular optimization used on data from a counting register of $N$ qubits has a smaller error bound (defined as $|\hat{\phi}-\phi|$) than a majority rule estimator for a counting register of $N+1$ qubits \cite{circular_optimization}. 

\subsection{Iterative Optimization}

The iterative PEA is another optimization that we tested. Unlike the circular optimization, the iterative optimization uses a different circuit than the base algorithm. It is so named because instead of performing the whole phase estimation in one circuit, it uses multiple smaller circuits to build up an estimation. In our implementation, we used three 3-qubit circuits. These circuits are identical except for the power of the controlled $U$ and the angle of the $R_z$ gate. A diagram for the circuits is pictured in Fig. \ref{fig:iterative}. Just as each counting qubit in the base circuit provides one binary digit of precision, each individual circuit in the iterative PEA provides one binary digit of precision. Instead of applying the controlled $U^{2^l}$ gate in parallel on one circuit, we create each PEA circuit with one counting qubit and apply a controlled $U^{2^l}$ gate, where $l$ is determined by the order in which the circuits are run. In place of the QFT, we perform a semiclassical QFT \cite{semiclassicalQFT}, which separates the typical QFT algorithm into a sequence of measurements and rotations requiring only classical connections between counting qubits. The iterative algorithm constructs $\hat{\phi}$ from the most ($(n-1)^{\text{st}}$) to least significant ($0^{\text{th}}$) bit with the following procedure, adapted from \cite{circular_optimization}:
\begin{figure}[tb]
    \centering
    \includegraphics[scale=0.60]{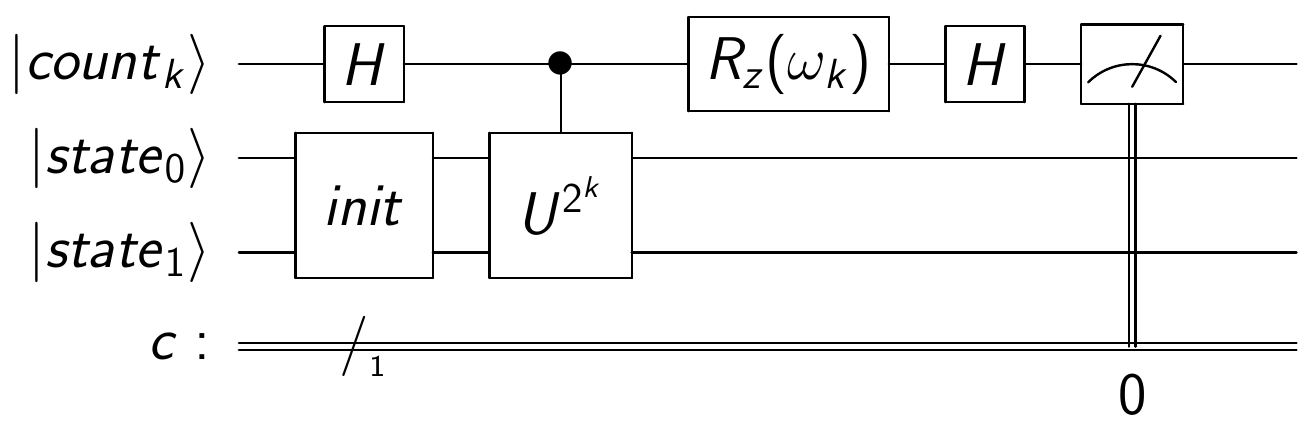}
    \caption[Quantum circuit diagram for the iterative optimization to the phase estimation algorithm]{Circuit diagram for the iterative optimization to the PEA with a Heisenberg Hamiltonian. The upper qubit forms the counting register and the lower two form the state register. The $R_z$ gate performs one step of the semiclassical QFT when parameterized by $\omega_k$ as defined in Eq. (\ref{equation:wk}). Note that for the iterative algorithm, three versions of this circuit must be run, each with different values for $k$, which affects both the $U^{2^k}$ gate and the $R_z$ gate.}
    \label{fig:iterative}
\end{figure}

\begin{enumerate}
    \item Run the $k=n-1$ circuit with $\omega_{n-1}=0$ to find the largest bit of $\hat{\phi}$.
    \item Repeat while $k>0$:
    \begin{enumerate}
        \item Decrement $k$.
        \item Calculate the rotation angle $\omega_k$ for circuit $k$ via the formula 
        \begin{equation}\label{equation:wk}
            \omega_k = -2\pi \sum_{j=0}^{n-k-2} \frac{b_{j+k+1}}{2^{j+2}},
        \end{equation}
        where $b_x$ refers to the measurement outcome of the $x^{th}$ circuit (the $x^{th}$ bit of $\hat{\phi}$). 
        \item Run circuit $k$ with the newly calculated $\omega_k$ to find $b_k$.
    \end{enumerate}
    \item Recover the estimator $\hat{\phi}$ from its binary expansion:
    \[
        \hat{\phi}_{iter} = \frac{1}{2^n}\sum_{x=0}^{n-1} 2^x b_x.
    \]
\end{enumerate}
In this way, we are able to perform each counting qubit of the PEA separately. To provide a fair comparison with the base algorithm, we used $n=3$ circuits. This optimization trades a larger multiplicity of circuits to run for smaller circuit sizes. It also has the same resolution as the base algorithm, permitting only values of $\hat{\phi}$ that are multiples of $1/2^n$, which is $1/8$ in our case. Note that circular statistics cannot be used to improve the iterative optimization because each bit is measured independently, whereas the base algorithm permitted us to determine coincidences among qubits.

\subsection{Bayesian Optimization}
Our final optimization for the PEA utilizes Bayesian statistics\index{Bayesian statistics}. In the versions of the algorithm we have already discussed, the estimator of the phase we are trying to determine is stored in the quantum state of a few qubits in some way. With the Bayesian PEA, we instead store the estimator as a parameter in the quantum circuit and use the states of the qubits to inform us how close our estimator is. The circuit is pictured in Fig. \ref{fig:bayes}. It is reminiscent of the circuits used for the iterative optimization, but the role of binary digit precision is replaced by a parameter $M$, which is repeatedly updated in the algorithm. It also includes an extra $R_z$ gate like the iterative algorithm; however, this $R_z$ gate serves a different purpose. In the Bayesian algorithm, this $R_z$ gate is also parameterized by $\theta$, which is our current guess for the true phase. When $\theta$ is close to the true phase $\phi$, the counting qubit ends up being almost purely in the $\ket{0}$ state; exactly how close depends on the parameter $M$. When $\theta$ is far from $\phi$ and $M$ is small, the counting qubit is almost purely in the $\ket{1}$ state. This property is used to guide our Bayesian updates.

Bayesian statistics is focused on the idea of updating a guess with newly acquired information. We begin with an initial guess of the mean and variance of the phase distribution, together known as the \textit{prior}. We choose a wide prior distribution to account for our total uncertainty about the phase. Specifically, we start with a normal distribution with a mean $\mu= \pi$ and variance $\sigma^2 = \pi^2$. Once an initial prior distribution is established, we perform the following algorithm, adapted from \cite{bayes1, bayes2, bayes3}:

\begin{figure}[tb]
    \centering
    \includegraphics[scale=0.6]{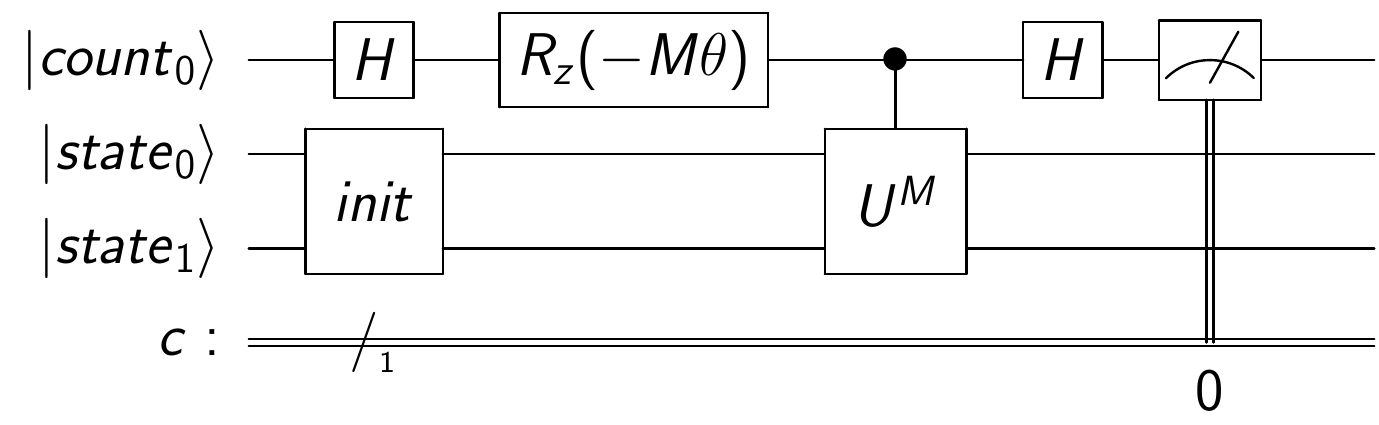}
    \caption[Quantum circuit diagram for the Bayesian optimization to the phase estimation algorithm]{Circuit diagram for the Bayesian PEA with a Heisenberg Hamiltonian. The upper qubit is the counting register and the lower two form the state register. The circuit is parameterized by $\theta$ and $M$, which approximately represent our estimate for the phase and the certainty of our estimate, respectively. In the Bayesian algorithm, several versions of this circuit are run as the parameters $M$ and $\theta$ are updated.}
    \label{fig:bayes}
\end{figure}

\begin{enumerate}
    \item Repeat $N$ times:
    \begin{enumerate}
        \item Sample $\theta$ from the current prior distribution and assign $M=\lceil 1.25/\sigma \rceil$.
        \item Run the quantum circuit in Fig. \ref{fig:bayes} with the parameters $M$ and $\theta$. Record the measurement result as $E \in \{0, 1\}$.
        \item Sample 100 $x$ values from the prior distribution. For each $x$ and based on the measurement result $E$, compute either
            \begin{equation}\label{equation:E0}
                P(E=0|x; \theta, M) = \frac{1}{2} + \frac{\cos(M(\theta - x))}{2}
            \end{equation}
            or
            \begin{equation}\label{equation:E1}
                P(E=1|x; \theta, M) = \frac{1}{2} - \frac{\cos(M(\theta - x))}{2},
            \end{equation}
        which tell us the likelihood of measuring 0 or 1 if the true phase $\phi$ was equal to $x$. 
        \item Generate 100 random values $u \in [0, 1)$ from a uniform distribution.
        \item For each $x_i$, if the result of Eq. (\ref{equation:E0}) or (\ref{equation:E1}) is greater than $u_i$, add $x_i$ to a new data set.
        \item Calculate the mean and variance of the new data set. Use the new mean and variance as the parameters of the updated prior distribution.
    \end{enumerate}
    \item Use the final distribution's mean $\mu_N$ as the approximator for $\phi$:
    \[
        \hat{\phi}_{Bayes} = \mu_N.
    \]
\end{enumerate}
~\\

The purpose of the Bayesian PEA is to perform successive updates on this prior in the hopes that it will become a sharp distribution around the true phase $\phi$, at which point sampling $\hat{\phi}$ from the distribution yields a good estimate. Because the phase estimate is stored in a classical parameter $\theta$ instead of the quantum state of a few qubits, we obtain a significantly better resolution for this algorithm than for the other optimizations, limited only by the numerical precision available to classical computers. The trade-off is the fact that the classical portion of the algorithm is probabilistic as well as the quantum portion, so our estimate may be incorrect even when no errors happen in the quantum calculation.

\section{Experimental Protocol}

For each algorithm, we replace the $U$ gate with the time evolution operator for the Heisenberg Hamiltonian as discussed in Section \ref{section:heisenberg}. After constructing the relevant circuits in Qiskit\index{Qiskit}, we run them with 8192 shots (the maximum allowed by IBM) to simulate evolution under the Hamiltonian for a time parameterized by the evolution parameter $\tau$. \index{Evolution parameter $\tau$} We choose $\tau$ to range from 0 to $2\pi$, representing one full period of time evolution\index{Quantum simulation}. At each $\tau$, the phase is estimated, allowing us to build up the phase as a function of $\tau$. This evolution is performed on both quantum simulators and real quantum computers provided by IBM Q. We then calculate the energy of the system with Eq. (\ref{eqn:energy}), using a finite difference method and taking the average derivative across the full evolution as our estimate.



\section{Results}
In this section we compare the results of simulations of quantum computers to the results of experiments on actual quantum computers. The former is useful because it provides an objective measure of the algorithm's individual performance, regardless of its implementation. The latter is important as it represents the current computational power available with modern quantum computers. 

\subsection{Simulation Results}
\begin{figure*}[tb]
    \centering
    \begin{subfigure}[t]{0.5\textwidth}
        \centering
        \includegraphics[scale=0.58]{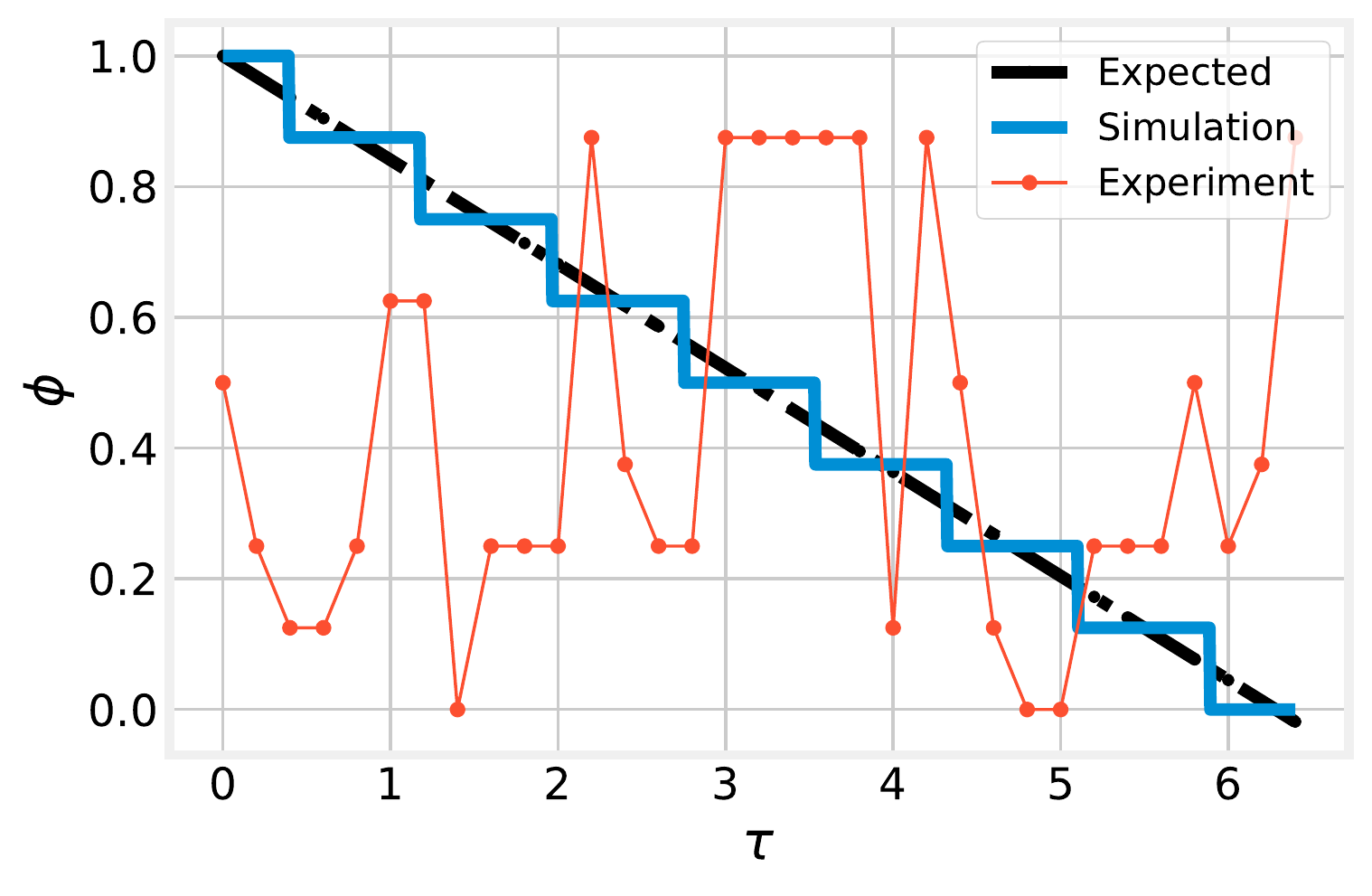}
        \caption{Base Algorithm}\label{fig:base_results}
    \end{subfigure}%
    \begin{subfigure}[t]{0.5\textwidth}
        \centering
        \includegraphics[scale=0.58]{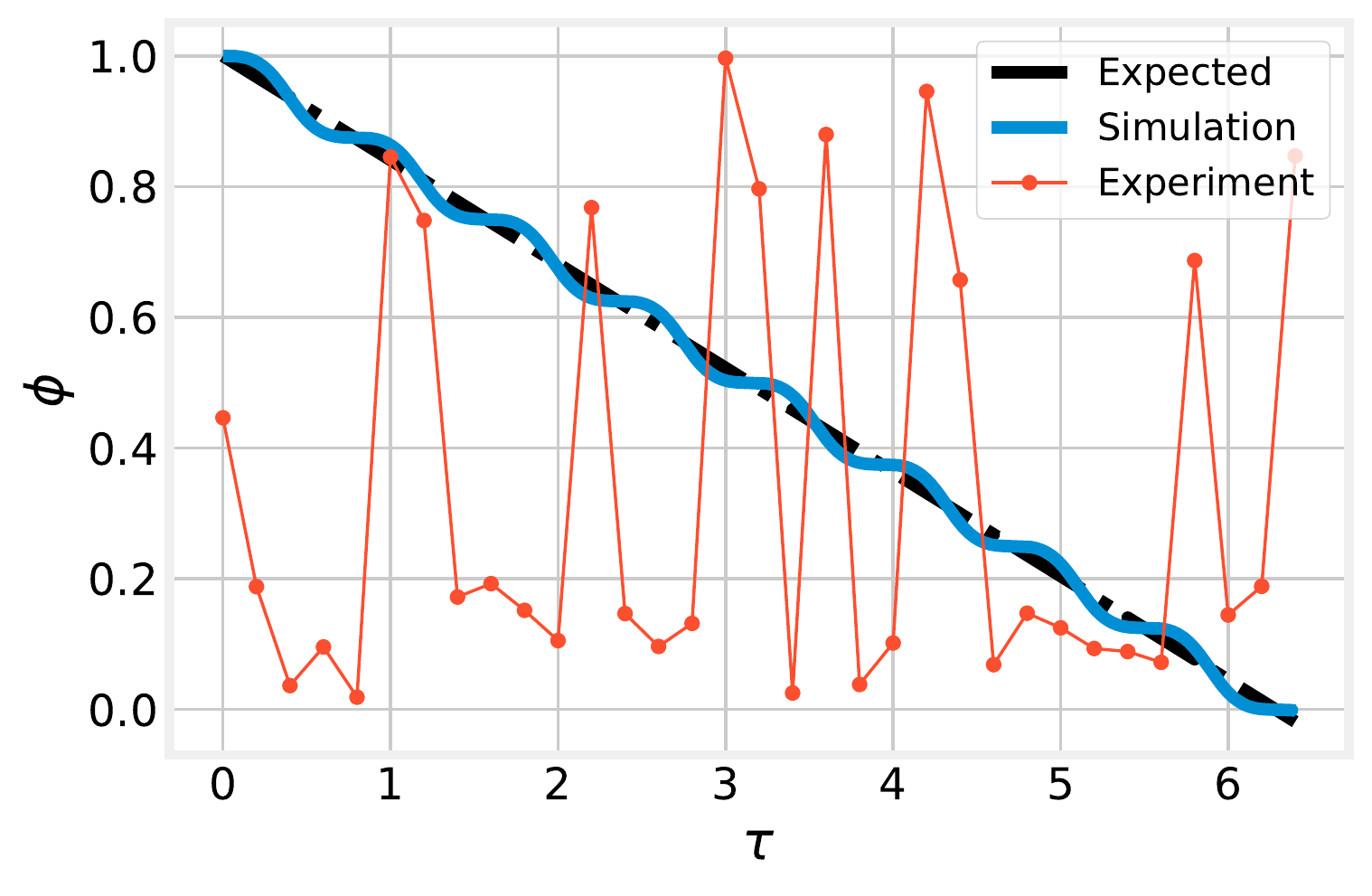}
        \caption{Circular Algorithm}\label{fig:circ_results}
    \end{subfigure}
    \begin{subfigure}[t]{0.5\textwidth}
        \centering
        \includegraphics[scale=0.58]{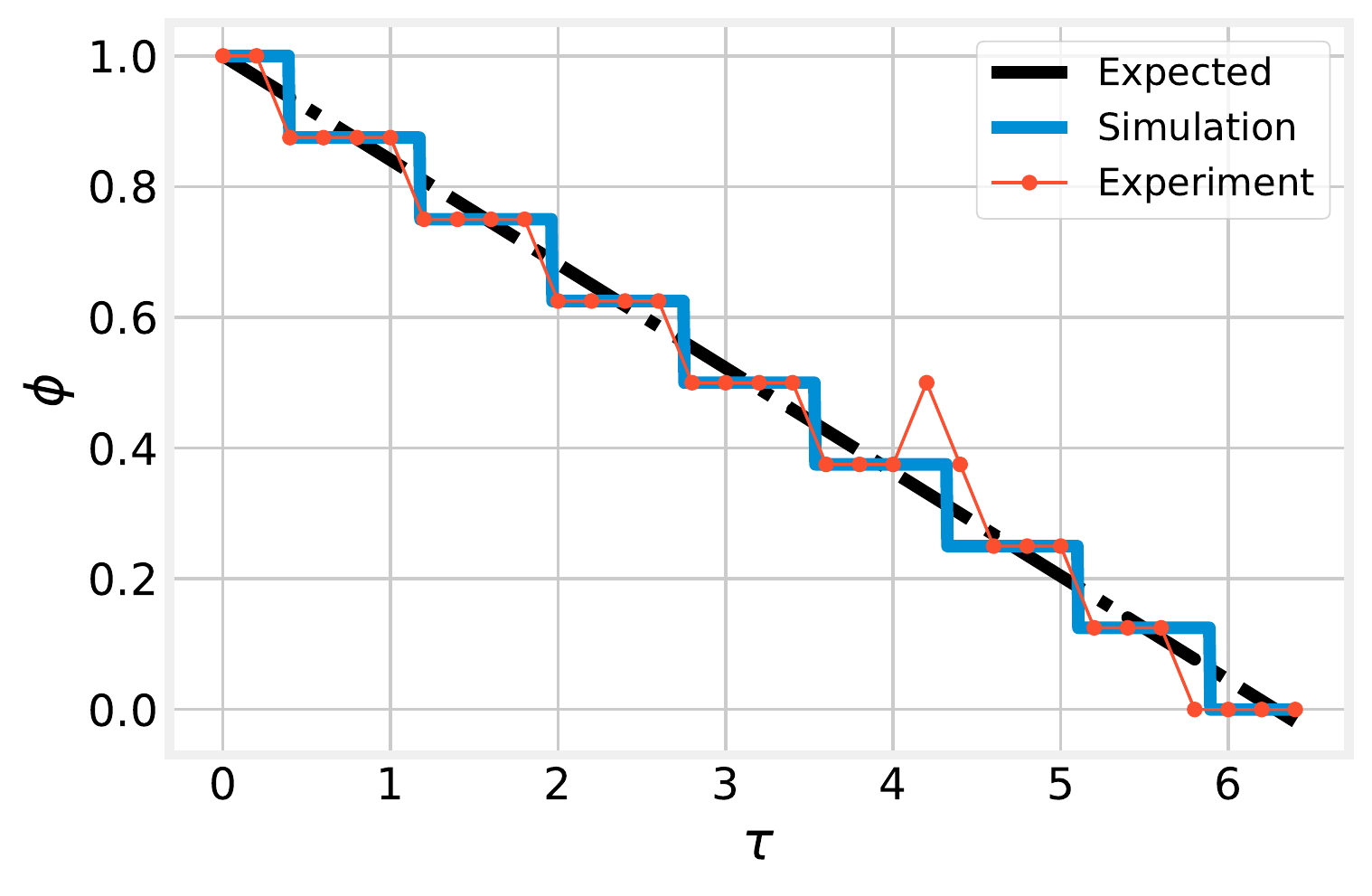}
        \caption{Iterative Algorithm}\label{fig:iter_results}
    \end{subfigure}%
    \begin{subfigure}[t]{0.5\textwidth}
        \centering
        \includegraphics[scale=0.58]{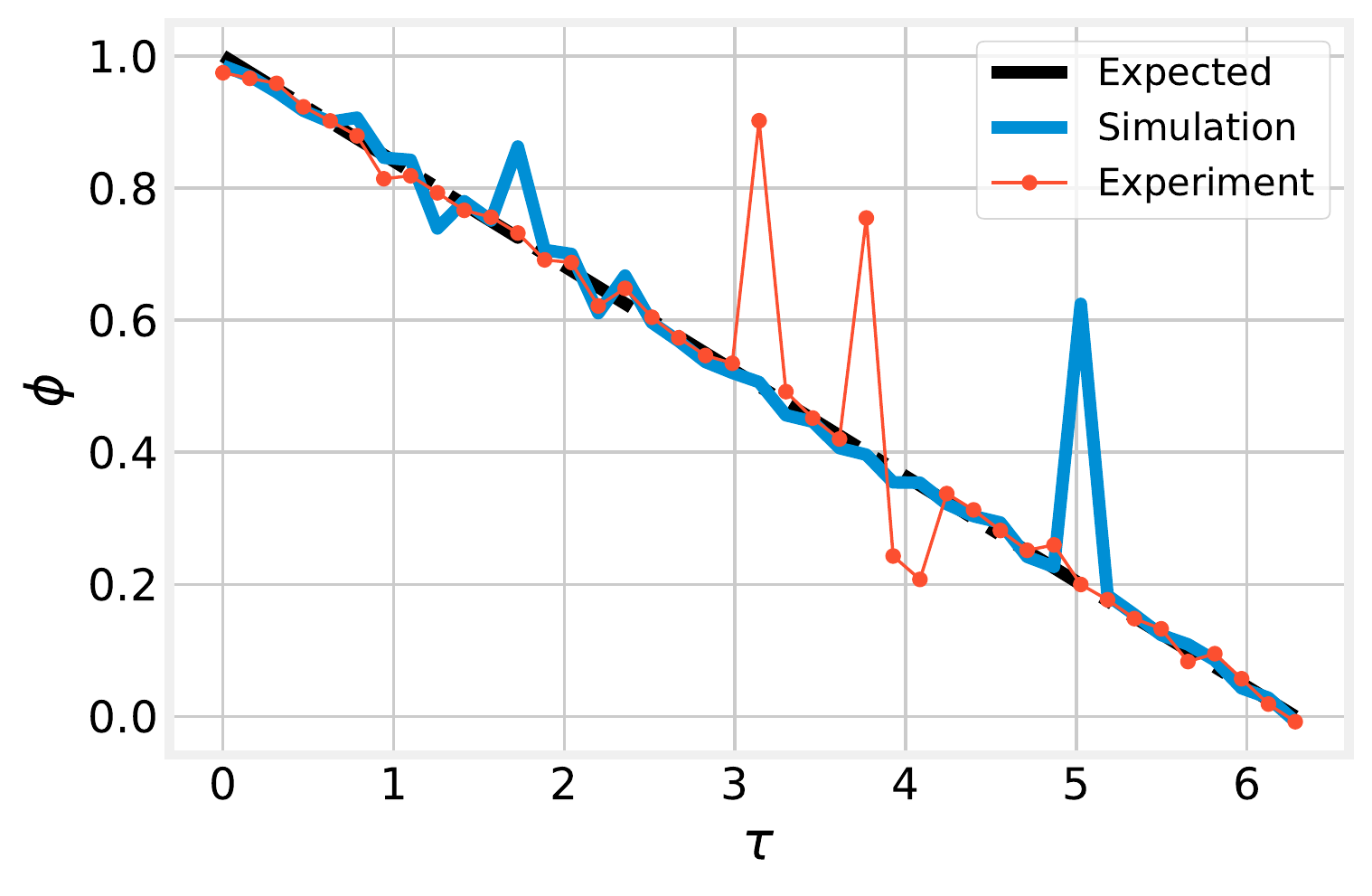}
        \caption{Bayesian Algorithm}\label{fig:bayes_results}
    \end{subfigure}
    \caption[Presentation of results from all four algorithms with expected, simulated, and experimental phase estimates]{Comparison of simulation and experimental data from each algorithm using a Heisenberg Hamiltonian with the parameter $\lambda = 4$, which gives an energy of $\epsilon=1$. All three lines plot the phase $\phi$ as a function of the evolution parameter $\tau$. Plotted in blue is the result of the evolution when the PEA is run on a simulation of a quantum computer and in orange is the result of the evolution when the PEA is run on an actual quantum computer at IBM. The actual phase of the Hamiltonian at each simulated time is plotted in dotted black lines.}
    \label{fig:all_results}
\end{figure*}

In simulations of noiseless quantum computers, we find that the circular optimization is ideal. A comparison of the four algorithms is presented in Fig. \ref{fig:all_results}, which plots calculated phase $\phi$ against evolution parameter $\tau$ for all four algorithms. In Figs. \ref{fig:base_results} and \ref{fig:iter_results} it is apparent that the simulation results of the base and iterative algorithms are identical. The flat plateaus followed by sharp jumps indicate the limited resolution of these two algorithms.

\onecolumngrid
\twocolumngrid

\begin{table}[tb]
\centering
\begin{tabular}{|c||c|c|c||c|c|c|}
\hline
\multirow{2}{*}{Algorithm} & \multicolumn{3}{c||}{Simulation} & \multicolumn{3}{c|}{Experiment} \\ \cline{2-7} 
                           & $\epsilon$ & \% Error & $R^2$  & $\epsilon$   & \% Error & $R^2$  \\ \hline \hline
Base                       &      1.013 & 1.3\%    & 0.986  &     0.887    &  11.3\%  & -1.245 \\ \hline
Circular                   &      1.009 & 0.9\%    & 0.997  &     0.419    &  58.1\%  & -1.140 \\ \hline
Iterative                  &      1.013 & 1.4\%    & 0.986  &     1.013    &  1.4\%   & 0.975  \\ \hline
Bayesian                   &      0.992 & 0.8\%    & 0.941  &     0.990    &  1.0\%   & 0.913  \\ \hline
\end{tabular}
\caption[Comparison of calculated energy values from simulations and experiments with error and correlation calculations]{Comparison of calculated energy values from simulations and experiments with each type of PEA. The energy $\epsilon$ is calculated from Eq. (2.4) based on the evolution results, with an expected value of $\epsilon = \lambda/4 = 1$. $R^2$ represents the coefficient of determination, calculated using the true phase function (plotted in black dotted lines in Fig. \ref{fig:all_results}) instead of a line of best fit.}
\label{table:results}
\end{table}

\noindent
By comparison, the continuous variation capability of the circular and Bayesian algorithms is visible in Figs. \ref{fig:circ_results} and \ref{fig:bayes_results}. The probabilistic nature of the Bayesian algorithm is apparent in the spikes away from the trend. All four algorithms' phase estimates follow the black dotted line in the figure, which indicates the true phase value. On the left side of Table \ref{table:results} we present the energy estimate $\epsilon$, the percent error from the true energy value, and the coefficient of determination $R^2$, which is calculated from the true phase plotted in black instead of a line of best fit. This $R^2$ value gives us quantitative measure of the reliability of our energy estimate. We can see that all four algorithms are good at estimating the energy of the system, boasting both a small percentage error and a high $R^2$ coefficient. Of particular note is the Bayesian algorithm, which obtains the closest energy estimate to the true value but has the worst $R^2$ value in simulations.

While the improved resolution of the circular and Bayesian algorithms makes the actual phase estimation more accurate, it does not significantly affect the average derivative in Eqn.~(\ref{eqn:energy}), from which the energy is calculated. This indicates that if an accurate result for one particular phase is desired, then the circular and Bayesian algorithms are ideal---but for energy estimation in simulation of time evolution, all four algorithms perform acceptably well in simulations.

\subsection{Quantum Experiment Results}
Looking at the orange lines in Fig. \ref{fig:all_results}, it is apparent that experiments on real quantum computers behave very differently from simulations of quantum computers. In simulations, there is no noise present; however on real quantum computers, noise enters as explained in Section \ref{sec:IBMQ}. The base and circular algorithms give phase estimates that vary wildly, almost never giving the true phase value. In fact, the noise seems to be sufficient to render the output of these algorithms essentially random. However, the iterative and Bayesian algorithm do not demonstrate the same result. In particular, we can see that the iterative algorithm's phase estimate in experiment only deviates from that of the simulation at three time steps, and the Bayesian algorithm's experimental phase estimate varies from the true value about as much as the simulation estimate does. In the right half of Table \ref{table:results}, we present parameters to quantify the results of the experiments. The base and circular algorithms demonstrate much worse performance, both having over 10\% error in the energy estimate. Even worse, their $R^2$ values are \textit{negative}. This is not a numerical error; instead, it indicates that the results of the experiment better approximate a horizontal line at the mean of the phase data than the expected phase value plotted in black in Fig. \ref{fig:all_results}. On the other hand, we also see that the iterative and Bayesian algorithms maintain a very small error in the energy calculation as well as a fairly high $R^2$ value, indicating that they perform very well in experiments on quantum computers. We can deduce that these latter two optimizations are essential to get reliable results from the PEA if it runs on a real quantum computer, regardless of whether phase estimates at particular times or average energy estimates are desired.

\section{Conclusions}
It is clear from the results above that the base and circular algorithms are not sufficiently robust (with respect to noise) to provide acceptable performance on real quantum computers. This is due to a combination of the significant amount of noise that quantum computers are subject to and of the length and size of the circuit---whereas the Bayesian and iterative algorithms use three-qubit circuits with no more than ten gates, the base and circular algorithms use a five-qubit circuit with over 50 gates. Both circuit size (qubit count) and circuit length (gate count) contribute to error\index{Error} buildup in quantum computers, though determining the relative contributions of each is outside of the scope of this paper.

We conclude that algorithms which recycle\index{Qubit!recycling} or otherwise reduce qubit count are likely to exhibit better performance than those which use more qubits. The iterative algorithm recycles a single qubit that is used with the functionality of multiple counting qubits. Although this was not possible in our implementation with the IBM Q system, it is possible for the state register to be re-used while the counting qubit is recycled \cite{photonic}. This produces the effect of reduced circuit size while maintaining identical functionality. Obtaining this effect in a quantum algorithm is typically difficult as larger circuit sizes generally provide a larger algorithmic space to work in, increasing the range of problems it can solve. The Bayesian algorithm confronts this challenge by effectively becoming a ``phase checker'' instead of a ``phase estimator,'' as it dynamically updates itself to produce an estimate that converges to the true phase. Circuit space is reduced by having the estimate stored as a real number in the classical system that controls the quantum computer instead of storing it on the computer itself.

\section{Discussion}

In the future, the PEA needs to perform its task accurately on large quantum systems in order to make quantum computing a useful computational tool in the hands of scientists. Our Hamiltonian\index{Heisenberg Hamiltonian} was a simple two-qubit, four-dimensional Hamiltonian whose eigenvalues are simple to calculate analytically, yet it is one of the larger Hamiltonians that is currently feasible to simulate without noise taking over\index{Noise}. For quantum simulation\index{Quantum simulation} to lead to novel scientific results, we need to be able to implement it on systems with Hamiltonians that are too large or complex to handle analytically or numerically. Examples include many-body problems, spin glasses, and open quantum systems \cite{quantum_simulation}. Our results indicate that the iterative and Bayesian optimizations should be used on these large Hamiltonians to be simulated accurately. Implicit in this problem is also the need to reduce a given Hamiltonian into a sequence of gates that can be implemented efficiently on a quantum computer as we did with the Heisenberg Hamiltonian in Section \ref{section:heisenberg}. Using the PEA to perform previously difficult tasks in cryptography or number theory, such as breaking RSA, requires it to be run on thousands to millions qubits. Considering we found that a quantum circuit of five qubits was too large to get reliable results from a real quantum computer, it is clear that the field has a lot of room to grow and develop.

While noise reduction is a common thread that underlines the development of quantum computers everywhere, it is also important to consider other paths to optimal quantum computer performance, such as qubit\index{Qubit} connectivity and error correction. Qubits on quantum computers are typically not fully connected, which complicates qubit allocation in algorithms and can result in the addition of unwanted yet necessary gates to quantum circuits \cite{transpiler}. Quantum error correction is another active field of research that aims to produce robust quantum computers that are subject to noise yet remain able to produce accurate results by recognizing when an error has occurred \cite{error_correction}. With such advancements in the field, we hope to utilize quantum technology to solve problems that have never before been solved.
\newpage
\begin{acknowledgements}
    We acknowledge support from the College of Physical and Mathematical Sciences at Brigham Young University. We also thank the members of the Quantum Information and Dynamics group at BYU for helpful and insightful discussions.
\end{acknowledgements}





 \bibliography{references}

\end{document}